\documentclass[10pt,preprint,tighten,flushrt]{aastex}

\newcommand{\kms}{\mbox{${\rm\,km\,s}^{-1}$}}
\newcommand{\kmspc}{\mbox{$\rm\,km\,s^{-1} pc^{-1}$}}

\newcommand{\msun}{\mbox{$\,M_\odot$}}

\newcommand{\beq}{\begin{equation}}
\newcommand{\eeq}{\end{equation}}
\newcommand{\hst}{{\it HST\/}}

\shorttitle{Eccentric Disks}
\shortauthors{Salow and Statler}

\begin{document}

\title{Eccentric Nuclear Disks with Self Gravity: Predictions for the 
Double Nucleus of M31}

\author{Robert M. Salow and Thomas S. Statler}
\affil{Department of Physics and Astronomy, 251B Clippinger Research
Laboratories, Ohio University, Athens, OH 45701, USA}

\slugcomment{\sl Accepted by The Astrophysical Journal Letters 2001 Feb 24}

\begin{abstract}
We present a method for constructing models of weakly self-gravitating,
finite dispersion eccentric stellar disks around central black holes.  The
disk is stationary in a frame rotating at a constant precession speed. 
The stars populate quasiperiodic orbits whose parents are numerically
integrated periodic orbits in the total potential.  We approximate the
quasiperiodic orbits by distributions of Kepler orbits dispersed in
eccentricity and orientation, using an approximate phase space
distribution function written in terms of the Kepler integrals of motion. 
We show an example of a model with properties similar to those of the
double nucleus of M31.  The properties of our models are
primarily determined by the behavior of the periodic orbits.  Self-gravity
in the disk causes these orbits to assume a characteristic radial
eccentricity profile, which gives rise to distinctive multi-peaked
line-of-sight velocity distributions (LOSVDs) along lines of sight near
the black hole.  The multi-peaked features should be observable in M31 at
the resolution of STIS.  These features provide the best means of
identifying an eccentric nuclear disk in M31, and can be used to constrain
the disk properties and black hole mass. 
\end{abstract}


\section{Introduction} \label{s.introduction}

{\it Hubble Space Telescope\/} (\hst) images have resolved the nucleus of
M31 into two distinct brightness peaks separated by $\sim 0\farcs 49$
(Lauer et al. 1993, 1998).  The optically fainter peak, P2, is located
near a UV-bright cluster, which is thought to be the true location of the
bulge photometric center and a supermassive black hole (BH; King et al.
1995; Kormendy \& Bender 1999, hereafter KB99).  The off-center peak, P1,
appears to be an intrinsic component of the nucleus, rather than an
intruding object (King et al. 1995, KB99).  The currently favored model
for the double nucleus (Tremaine 1995, hereafter T95) consists of an
eccentric nuclear disk of stars on apse-aligned Kepler orbits (see
Emsellem \& Combes 1997 for an alternative).  P1 represents the part of
the disk near apocenter, where slower moving stars accumulate. 

T95's disk is represented by three aligned Keplerian ringlets with
outwardly decreasing eccentricities.  New data from the
Canada-France-Hawaii Telescope (CFHT) Subarcsecond Imaging Spectrograph
(SIS) (KB99) are consistent with predicted asymmetries in the T95 model,
including the offset rotation center, the asymmetric rotation amplitudes,
the low dispersion at P1, and the dispersion peak near P2.  The \hst\
Faint Object Camera (FOC) observations of Statler et al. (1999, hereafter
SKCJ) resolve the rotation curve and verify the existence of the
dispersion peak, which has $\sim 400 \kms$ amplitude at FOC resolution. 
One-dimensional kinematic profiles from Space Telescope Imaging
Spectrograph (STIS) spectra and two-dimensional data from the OASIS
instrument on the CFHT (Bacon et al. 2001, hereafter B01) are both
consistent with the FOC kinematics and the eccentric disk picture, when
the data sets are spatially registered. 

T95's conceptual model needs to be extended to include self-gravity and
finite velocity dispersion.  Sridhar \& Touma (1999, hereafter ST99)
explore orbits in Kepler potentials with lopsided perturbations and find a
class of periodic loop orbits elongated in the same sense as the
perturbation.  They point out that the nearly elliptical periodic parents
of such orbits can be seen as the backbone of an eccentric disk with
self-gravity and finite dispersion.  Statler (1999, hereafter S99)
computes periodic loop orbits for a uniformly precessing Tremaine-like
disk model.  He predicts a characteristic non-monotonic radial
eccentricity distribution for the periodic orbits, resulting in a disk
structure significantly different from that of the original T95 model. 
B01 present numerical simulations of an $m=1$ mode in a cold disk with a
central BH.  They find a similar eccentricity structure to that
predicted by S99.  Tremaine (2001) studies slow $m=1$ modes in
axisymmetric nearly Keplerian disks using linear perturbation theory.  He
finds stable modes which have a different eccentricity structure from that
seen by either S99 or B01. 

In this {\em Letter\/}, we construct approximate self-consistent eccentric
disk models with self-gravity and finite velocity dispersion.  The details
of model construction are described in \S\ \ref{s.disks}.  We then briefly
discuss the properties of such models in \S\ \ref{s.models}.  We show in
\S\ \ref{s.losvd} that the line of sight velocity distributions (LOSVDs)
on particular lines of sight have unusual bimodal shapes, and can be used
to identify the presence of an eccentric disk in M31.  Finally, \S\
\ref{s.discussion} discusses the connection with other work. 

\section{Model Construction} \label{s.disks}

We postulate a disk whose density distribution is fixed in a frame
rotating at constant angular speed $\Omega$ about a BH of mass
$M_\bullet$.  The disk is built from quasiperiodic loops (ST99) that
librate about periodic parents. The latter are nearly Kepler ellipses in
the rotating frame for small disk masses.  We approximate the
quasiperiodic orbits about a sequence of parents by a distribution of
Kepler orbits dispersed in eccentricity and orientation, using the
distribution function (DF)
\begin{equation}
f(a,e,\gamma) = F(a)\exp \left\{ - {[e-e_0(a)]^2 \over 2 {\sigma_e}^2} 
\right\}\exp \left[ - {\gamma^2 \over 2 {\sigma_\gamma}^2} \right].
\label{e.df}
\end{equation}
In equation (\ref{e.df}) the semimajor axis $a$, the eccentricity $e$, and
$\gamma$, the angle between the disk major axis and the Runge-Lenz vector
$\vec{A}=\vec{v}\times\vec{h}-GM\hat{r}$, are integrals of motion in the
unperturbed potential.  The function $e_0(a)$ describes the sequence of
parent orbits that form the backbone of the disk.  The constants
$\sigma_e$ and $\sigma_\gamma$ are the dispersions in $e$ and $\gamma$. 
The function $F(a)$, giving the mass per unit interval of semimajor axis,
controls the radial mass distribution.  We adopt
\begin{equation}
F(a) = \mbox{\rm max}(a-1,0)\exp \left[ - {(a-a_0)^2 \over 2 {\sigma_a}^2} 
\right],
\label{e.fofa}
\end{equation}
where the constant $a_0$ sets the length scale for the model, and
$\sigma_a$ is the width of the density distribution.  This $F(a)$ is
chosen to produce density figures with limited radial extent and strong
central density minima. It is not intended to fit in detail the surface
photometry of the M31 nucleus, where the putative disk is not radially
truncated. Nonetheless, the models do reproduce many of the observed
characteristics of the inner regions, the predicted LOSVDs at small radii
being essentially independent of the outer disk structure. 

Starting from an initial guess for $e_0(a)$, the DF is written in terms of
position and velocity using the Keplerian relations for $a$, $e$, and
$\gamma$, and integrated over velocity to obtain the density, normalized
for a total disk mass $m$. The potential is found using a standard Fourier
method on a cartesian grid.  The disk potential is added to that of the
central BH, and the combined potential is rotated at frequency
$\Omega$ about the center of mass. We then numerically integrate to find a
new set of periodic parent orbits, which become the backbone for the next
iteration.  Iterations are terminated when the fractional change in the
density per iteration is $<5\%$ everywhere and $<1\%$ on average.  The
results are not sensitive to the initial $e_0(a)$.

\begin{figure}[t]
\plotone{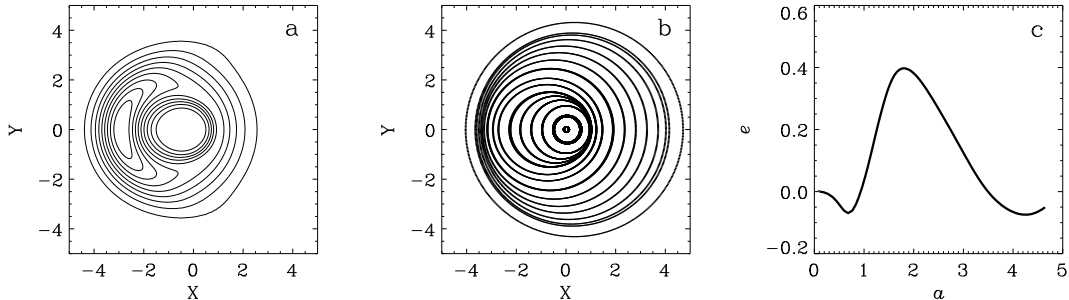}
\caption{\footnotesize
($a$) Density contours for the fiducial model.  Contours are at 0.1, 0.2,
..., 1.0 of the maximum density.  The central point mass is at (0,0), near
the point of minimum density.  ($b$) Uniformly precessing periodic orbits
in the total potential.  The radial variation of eccentricity is a
consequence of disk self-gravity. ($c$) Eccentricity of the orbits in
($b$) plotted against semi-major axis;  this is the function $e_0(a)$ in
equation (\ref{e.df}). \label{f.construction}}
\end{figure}

\section{Model Properties} \label{s.models}

A model is specified by the parameters $\epsilon=m/M_\bullet$, $\Omega$,
$\sigma_a$, $\sigma_e$, and $\sigma_\gamma$.  Models are computed in
dimensionless units where $G=M_\bullet=1$ and $a_0=2$.  Figures 1, 2, and
3 show results for a fiducial model with $\epsilon=0.03$, $\Omega=0.015$,
$\sigma_a=1.0$, $\sigma_e=0.175$, and $\sigma_\gamma=0.4$. 

Figure 1a shows a contour plot of the surface density.  The disk is
non-axisymmetric, with a density maximum near $x=-3$.  The outer cutoff
reflects the adopted form of $F(a)$.  Figure 1b shows the set of periodic
parent orbits.  These orbits have the non-monotonic run of eccentricity
with semimajor axis argued for by S99 (his Fig. 3).  The sequence is
described by the function $e_0(a)$, plotted in Figure 1c.  Notice that
$e_0(a)$ peaks near $a=a_0$, and becomes negative at larger $a$;  orbits
with negative eccentricities have apocenters at $x>0$.  The basic shape of
this function is characteristic of all the models we have computed. 

\begin{figure}[t]
\plotone{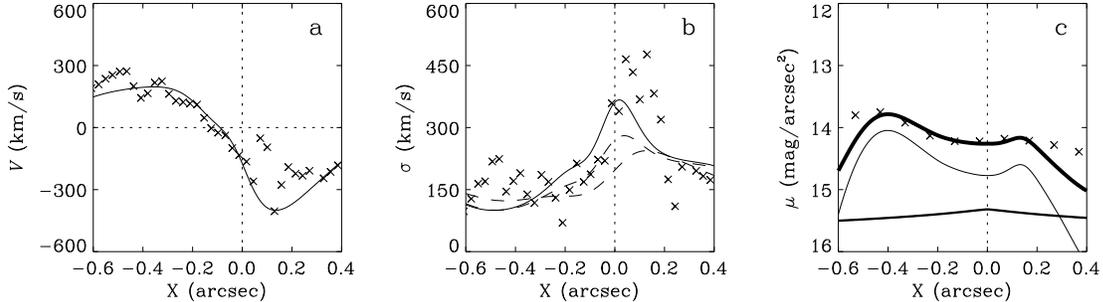}
\caption{\footnotesize
Major axis kinematic and photometric profiles for the fiducial model,
scaled to M31.  The added bulge model has $\eta=2.28$ and scale length
$r_0=85 \arcsec$, scaled to resemble B01's MGE model from $4\arcsec$ to
$10\arcsec$.  Disk and bulge have $V$-band $M/L=4$ (solar units, T95), and
$M_\bullet = 3.2\times 10^7 \msun$.  The length scale is set by the
$0\farcs 44$ separation between P1 and the UV peak (B01).  ($a$) Rotation
curve and ($b$) Velocity dispersion convolved to the resolution of FOC
({\em solid line\/}).  Crosses show FOC profiles;  error bars are omitted
for clarity.  Dashed lines in ($b$) show profiles for $0\farcs 1$ and
$0\farcs 35$ slit widths.  ($c$) Surface brightness profiles for the disk
({\em thin line\/}), bulge ({\em medium line\/}), and the total ({\em
heavy line\/}) integrated over a $0\farcs 35$ wide slice, with \hst\
$I$-band data ({\em crosses\/}; from Fig.\ 8 of KB99, $14.4 \,{\rm
mag}\,{\rm arcsec}^{-2}$ zero point, includes $12.56$, $1.37$, and $0.47
\,{\rm mag}\,{\rm arcsec}^{-2}$ shifts for the $I$-band zero point (J.
Kormendy, private communication), $V$-$I$, and compensation for the
model's truncation within the $0\farcs 35$ slice, respectively.
\label{f.observations}}
\end{figure}

Observable velocity moments can be obtained by integrating the DF in
equation (\ref{e.df}).  Projected profiles on the disk major axis are
shown in Figure 2.  For the sake of realism we have scaled the model to
the parameters of M31, and simulated the FOC observations.  The bulge and
central cusp are approximated by a single spherical, non-rotating
$\eta$-model (Tremaine et al. 1994) that includes the influence of the
BH.\footnote{The effect of the bulge potential on the disk is ignored; the
bulge-induced precession frequencies of Kepler orbits in the absence of
the disk potential are $<0.001$.}\ The disk is inclined at $77 \arcdeg$,
and the line of sight is perpendicular to the major axis.  Moderate
rotations of the model in the disk plane affect the results only slightly. 
The observables are integrated over the FOC slit as projected onto the
disk plane. 

The rotation curve and dispersion profile are shown as the solid lines in
Figures 2a and 2b.  Crosses show the FOC data for comparison; both
profiles have been shifted by $+0\farcs 025$ spatially and the rotation
curve has been shifted upward by $30 \kms$, as recommended by B01. Figure
2c shows the surface brightness profiles for the disk, bulge, and the sum,
compared with $I$-band \hst\ WFPC2 photometry.  This model is able to
reproduce many of the salient features in the observed profiles, even
though it is not a rigorous fit.\footnote{Because the disk models are
radially truncated, the fit to the data for $x<-0\farcs 6$ and $x>0\farcs
2$ is poor.}\ These features include the relative brightnesses of P1 and
P2, the shape of the rotation curve, the offset of the rotation center,
and the low dispersion of P1.  A dispersion spike appears near P2 in both
model and data.  Its detailed properties are problematic, and will be
discussed in \S\ \ref{s.discussion}. 

Changing the parameters from their fiducial values alters the disk
properties.  Axisymmetry decreases monotonically with increasing
$epsilon/\Omega$.  From the ratio of the projected surface brightness
peaks, we find that disks become axisymmetric to within $5\%$ for
$\epsilon/\Omega<0.6$.  M31-like models require $\epsilon/\Omega>1$. We
find converged models only for $0.5 \lesssim \epsilon/\Omega \lesssim 2.0$
and $\epsilon \lesssim 0.21$. 

Disks are also made more axisymmetric by increasing $\sigma_\gamma$, which
spreads the orientations of the dispersed ellipses.  Increasing
$\sigma_e$, on the other hand, makes the disk more lopsided, essentially
because the contrast in linear density on a single Kepler orbit,
$\rho_{apo}/\rho_{peri} = (1+e)/(1-e)$, is a concave-up function of $e$. 
We find empirically that we need $\sigma_e \geq 0.15$ and $\sigma_\gamma
\leq 0.6$ to produce M31-like models with large eccentric density peaks. 
Finally, the effect of increasing $\sigma_a$ is to broaden the mass
distribution radially, while having minimal effect on the kinematics or
axisymmetry. 

\section{LOSVDs} \label{s.losvd}

All of our models share the same characteristic eccentricity structure for
the parent periodic orbits (Fig.\ \ref{f.construction}b,c).  The
eccentricity structure, modulated by the velocity dispersion, determines
the LOSVDs.  We find that the LOSVDs along certain lines of sight hold the
key for identifying disks of this type and for measuring their basic
parameters. 

\begin{figure}[t]
\plotone{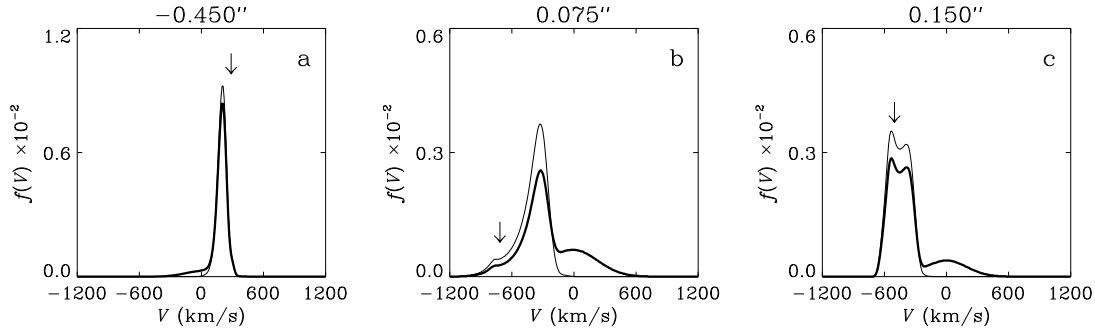}
\caption{\footnotesize
LOSVDs for the M31-scaled fiducial model, integrated over a $0\farcs 1$
wide slit and convolved to STIS resolution.  The {\em thin\/} ({\em
thick\/}) line shows the LOSVD {\em without\/} ({\em with\/}) the bulge
contribution.  Arrows mark the circular speed at the tangent point.  ($a$)
LOSVD through P1, $x=-0\farcs 450$ from the UV peak, similar to that
expected for a cold, axisymmetric disk with outwardly decreasing density. 
($b$) LOSVD through P2, $x=0\farcs 075$ from the UV peak.  The peak at
supra-circular velocity results from the eccentricity structure, the large
peak from the outwardly increasing density.  ($c$) LOSVD $0\farcs 075$
past P2, $x=0\farcs 150$ from the UV peak.  The density decreases
outward; the smaller peak is from the eccentricity gradient.
\label{f.losvds}}
\end{figure}

In Figure 3 we show three LOSVDs for the fiducial model, oriented as in
Figure 2 and scaled to the parameters of M31.  The thick and thin lines
show the LOSVDs with and without the bulge model described in \S\
\ref{s.models}.  We approximate the bulge LOSVD by a Gaussian with the
correct projected dispersion of the adopted $\eta$-model, ignoring the
higher moments. The LOSVDs are convolved to approximately the spatial and
spectral resolution of the STIS data, assuming Gaussian point- and
line-spread functions. 

Figure 3a shows the LOSVD for a line of sight through P1.  The shape is
similar to that expected for a cold, axisymmetric disk.  The LOSVD peaks
near the tangent point velocity, and shows an extended tail arising from
material in front or behind the tangent point.  The tail in ordinary disks
is weak when the density decreases outward.  The model LOSVD has this
form; the bulge contribution is minimal, due to the large disk density at
P1. 

Figure 3b shows the LOSVD for a line of sight through P2, $0\farcs 075$ on
the opposite side of the central mass. Here the disk LOSVD is double
peaked.  The peak at large negative velocities corresponds to the tangent
point, which falls near the pericenters of orbits with substantial
eccentricities (Fig.\ \ref{f.construction}b), creating a peak at
supra-circular velocities.  The peak at smaller negative velocities has
the same origin as the low velocity tail in P1, but the slope is reversed
for two reasons: first, the line of sight traverses the disk's central
hole, where the density increases outward; second, the outward decrease in
eccentricity amplifies the line of sight velocity of material in front or
behind the tangent point.  The former effect dominates along this line of
sight, so the location of the low speed cutoff is a function of the slit
width. 

Figure 3c shows the LOSVD for a line of sight an additional $0\farcs 075$
past P2. The LOSVD changes significantly over very small spatial scales,
owing both to the near Keplerian potential and the negative gradient in
$e_0(a)$. The tangent point peak has shifted by $\sim 250\kms$ over
$0\farcs 075$.  Roughly 3/4 of this shift results from the Keplerian drop
in circular speed, and the remaining 1/4 from the eccentricity gradient. 
The peak at slower speeds results entirely from the eccentricity gradient,
since the line of sight does not traverse the central hole. 

All of our disk models show bimodal LOSVDs in projection near the
secondary density peak. {\em These LOSVDs are generic to the models,
arising from the density and eccentricity structure, and should be
observable at STIS resolution.} They therefore provide the best means of
unambiguously identifying an eccentric nuclear disk in M31. 

\section{Discussion} \label{s.discussion}

We have presented a method for constructing approximate models of
near-Keplerian eccentric disks with self gravity and finite velocity
dispersion, appropriate to the M31 double nucleus.\footnote{The model does
not show a ``P1 wiggle'' in the rotation curve, which S99 cites as a
possible signature of self-gravity. Contrary to S99's speculation, this
feature is washed out in all but the coldest models. Since the P1 wiggle
is not seen in the STIS rotation curve either, we conclude that it is
probably an artifact of the FOC data.}\ The properties of the models are
dictated by the ``backbone'' periodic orbits, which show a characteristic
radial profile (S99), giving rise to distinctive LOSVDs for lines of sight
through the secondary density maximum, near the central mass. 

The details of the dispersion peak (Figure 2b) deserve special mention. 
Its amplitude can be tuned by changing $M_\bullet$ or other model
parameters.  The observed maximum, however, is offset from the UV cluster
by a distance that depends on resolution, increasing from $0\farcs 1$ for
the FOC data to $0\farcs 2$ for the OASIS data (B01).  We find that, while
the model peak also shifts away from the origin with widening slit width
(Fig.\ 2b, dashed lines), the offset is smaller by $\sim 0\farcs 06$.  It
is difficult to move the dispersion maximum away from the position of the
BH.  The bulge dispersion must peak at the origin if the BH dominates the
gravity.  Either this peak will shine through the density minimum in the
disk, or disk material orbiting at very small radii will produce the same
effect.  We see two possible sources for the discrepancy in position: 
first, that the models are incomplete, missing some essential bit of
physics; second, that the blue cluster is a red herring, not marking the
true location of the BH. 

Our fiducial model is similar to B01's $N$-body simulation of a strong
$m=1$ mode in a cold, thin disk with a central BH.  The peak locus in
their particle density plot in the $(e,a)$ plane is similar to our
function $e_0(a)$ (Fig.\ 1c).  B01 note in particular that the pericenters
of orbits in the inner and outer parts of the disk are in phase
opposition, as predicted by S99. They find a pattern speed of $\Omega=3
\kmspc$, compared to our fiducial model with $\Omega_p=13.6 \kmspc$.  On
the other hand, Sambhus and Sridhar (2000), using a variant on the
Tremaine \& Weinberg (1984) method, obtain $20 \pm 12 \kmspc$ and $34 \pm
8 \kmspc$ from two different fits to the bulge.  Thus neither data nor
models at present provide good constraints on the pattern speed. 

How such a large $m=1$ mode could arise remains problematic. B01 indicate
that the nonlinear modes seen in their simulations can grow spontaneously
from axisymmetric initial conditions. On the other hand, Tremaine (2001)
examines slowly precessing $m=1$ modes in nearly Keplerian disks,
concluding that all such disturbances are stable and tend not to have the
eccentricity structure shared by our models and B01's simulation. 

The true test of the eccentric disk picture for the M31 nucleus lies in
the line of sight velocity distributions near P2. Our models predict that
the LOSVDs in this region should be distinctive and probably bi- or even
tri-modal.  The widths of the peaks, and how they shift in velocity and
vary in amplitude as a function of position, will constrain the disk's
velocity dispersion and its density and eccentricity structure. 
Extracting the full LOSVD from the M31 spectra demands much higher
signal-to-noise ($S/N$) than obtaining $V$ and $\sigma$ alone.  B01 defer
discussion of the higher moments in the STIS data because of $S/N$ issues,
though preliminary results look intriguing (E. Emsellem, private
communication). 

\acknowledgments

This work was supported by NSF CAREER grant AST 97-03036.  We thank Scott
Tremaine for providing code to compute self-consistent $\eta$ models, and
John Kormendy, Ralf Bender, and the referee, Eric Emsellem, for helpful 
comments.


\begin{references}

\reference{Bac01} Bacon, R., Emsellem, E., Combes, F., Copin, Y., Monnet,
	G., \& Martin, P.\ 2001, \aap, submitted; astro-ph/0010567 (B01)
\reference{EmC97} Emsellem, E. \& Combes, F. 1997, \aap, 323, 674
\reference{KSC95} King, I. R., Stanford, S. A., \& Crane, P. 1995, \aj,
	109, 164
\reference{KoB99} Kormendy, J., \& Bender, R. 1999, \apj, 522, 772 (KB99)
\reference{Lau93} Lauer, T. R., Faber, S. M., Groth, E. J., Shaya, E. J.,
	Campbell, B., Code, A., Currie, D. G., Baum, W. A., Ewald, S. P.,
	Hester, J. J., Holtzman, J. A., Kristian, J., Light, R. M., \&
	Westphal, J. A. 1993, \aj, 106, 1436
\reference{Lau98} Lauer, T. R., Faber, S. M., Ajhar, E. A.,
	Grillmair, C. J., \& Scowen, P. A. 1998, \aj, 116, 2263
\reference{Sam00} Sambhus, N. \& Sridhar, S. 2000, \apjl, 539, L17
\reference{Sri99} Sridhar, S. \& Touma, J. 1999, \mnras, 303, 483 (ST99)
\reference{Sta99} Statler, T.\ S.\ 1999, \apjl, 524, L87 (S99)
\reference{SKCJ99} Statler, T.\ S., King, I.\ R., Crane, P., \& Jedrzejewski,
	R.\ I.\ 1999, \aj, 117, 894 (SKCJ)
\reference{Tre94} Tremaine, S., Richstone, D.\ O., Byun, Y.-I., Dressler,
	A., Faber, S.\ M., Grillmair, C., Kormendy, J., \& Lauer, T.\
	R.\ 1994, \aj, 107, 634
\reference{Tre95} Tremaine, S. 1995, \aj, 110, 628 (T95)
\reference{Tre01} Tremaine, S. 2001, \aj, in press (astro-ph/0011571)
\reference{TrW84} Tremaine, S.\ \& Weinberg, M.\ D.\ 1984, ApJ, 282, L5

\end{references}
\end{document}